 \documentclass[final,5p,times,twocolumn]{elsarticle}

\usepackage{amssymb}

\usepackage{balance}

\usepackage[usenames]{color}

\usepackage{lipsum}
\makeatletter
\def\ps@pprintTitle{%
 \let\@oddhead\@empty
 \let\@evenhead\@empty
 \def\@oddfoot{}%
 \let\@evenfoot\@oddfoot}
\makeatother

\begin{document}

\begin{frontmatter}

\title{\vspace*{0.5cm}Electronic and thermal conduction properties of halogenated \\porous graphene nanoribbons}

\author[1,2]{G. A. Nemnes\corref{mycorrespondingauthor}}
\cortext[mycorrespondingauthor]{Corresponding author. Tel.: +40 (0)21 457 4949/157. \\ {\it E-mail address:} nemnes@solid.fizica.unibuc.ro (G.A. Nemnes).}
\address[1]{University of Bucharest, Faculty of Physics, Materials and Devices for Electronics and Optoelectronics Research Center,\\ 077125 Magurele-Ilfov, Romania}
\address[2]{Horia Hulubei National Institute for Physics and Nuclear Engineering, 077126 Magurele-Ilfov, Romania}
\address[3]{School of Science and Engineering, Reykjavik University, Menntavegur 1, IS-101 Reykjavik, Iceland}

\author[2]{Camelia Visan}

\author[3]{A. Manolescu}





\begin{abstract}
In the framework of density functional theory (DFT) calculations we investigate the electronic and thermal properties of porous graphene (PG) structures passivated with halogen atoms as possible candidates for efficient thermoelectric devices. Armchair and zig-zag halogenated PG nanoribbons are analyzed comparatively. The electronic properties are consistent with the expected behavior for the two types of terminations, however with marked influences introduced by the different halogen atoms. Depending on the pore sizes and halogen type pseudo-gaps in the phononic band structure are visible in the low frequency range, which are particularly important for the thermal conduction at low temperatures. The gaps are systematically displaced towards lower energies as the atomic number of the halogen increases. At the same time, the electronic gap decreases, which is also essential for attaining a large figure of merit in a thermoelectric device. This indicates the possibility of tuning both electronic and thermal properties of PG structures by halogen passivation. 
\end{abstract}


\end{frontmatter}


\section{Introduction}

Graphene is a unique material with distinguished physical properties, amongst which are the remarkably high electric and thermal conductivities. The high mobility of graphene sheets is exploited in a vast number of applications, such as field effect transistors working in the high GHz regime \cite{schwierz}, while the large optical transparency makes them candidates for transparent conducting electrodes \cite{park}. 

Thermoelectric applications based on graphene have been also envisioned, although so far limited in number. This is due to the two major shortcomings that should be taken care of, namely the large thermal conductivity and a rather small Seebeck coefficient \cite{xu} of pristine graphene. Nanostructuring of the graphene sheets constitutes a feasible route for increasing the thermoelectric figure of merit and a giant thermoelectric effect was indicated \cite{dragoman}. An enhancement of the thermoelectric properties was also reported by introducing boron nitride domains and doping in graphene \cite{liu,vishkayi,mandal}.

For lowering the thermal conductivity several approaches have been considered. In disordered edge structures \cite{sevincli} or carbon nanotube networks \cite{gupta} the phonon scattering may be enhanced. Other approaches include defect engineering \cite{haskins}, edge passivation \cite{hu} and more recently nanopore patterning \cite{haskins,chang,sadeghi,chang2,hossain}. In addition, two-dimensional carbon allotropes \cite{nemnes} may provide alternative routes to graphene.  

Porous graphene nanostructures found applications in a large variety of nanodevices and processes, such as DNA sequencing \cite{heerema}, gas purification \cite{hauser}, hydrogen storage \cite{reunchan,ao}, electrochemical energy storage \cite{han} and supercapacitors \cite{fan,zhang}. Synthesis pathways, properties and potential applications of porous graphene are presented in recent reviews \cite{pengtao,russo,jiang}.

With the ongoing development of state-of-the-art fabrication techniques, porous graphene membranes
are in the focus of recent investigations. By employing focused ion beam (FIB), drilled apertures on the freestanding graphene have been experimentally produced \cite{liu2}, which influence both electrical and thermal conduction. In particular, periodic arrangements of the pores may introduce phononic band gaps. Systems of this kind, known as phononic crystals, received a lot of attention lately \cite{zen}. Adjusting the phononic band gap has become one important goal for the design of new acoustic wave devices, but also for the next generation thermoelectric devices, which require high electrical and low thermal conductivities. 

Halogen-terminated graphene nano-flakes \cite{tachikawa} and graphene based structures with absorbed halogen \cite{xu2} have been reported to possess tunable electronic properties.  
Halogenated graphenes were produced by thermal exfoliation of graphite oxide in different gaseous halogen atmospheres \cite{poh}. The structures analyzed by Raman spectroscopy show specific differences due to halogen functionalization. Moreover, electrochemical properties such as the peak-to-peak separation in the cyclic voltametric experiments, heterogeneous electron transfer rates and the charge-transfer resistance are shown to be directly linked to the atomic number of the halogen.

\begin{figure}[t]
\centering
\hspace*{-6cm}\vspace*{0.0cm}APG1\\
\includegraphics[width=3.5cm]{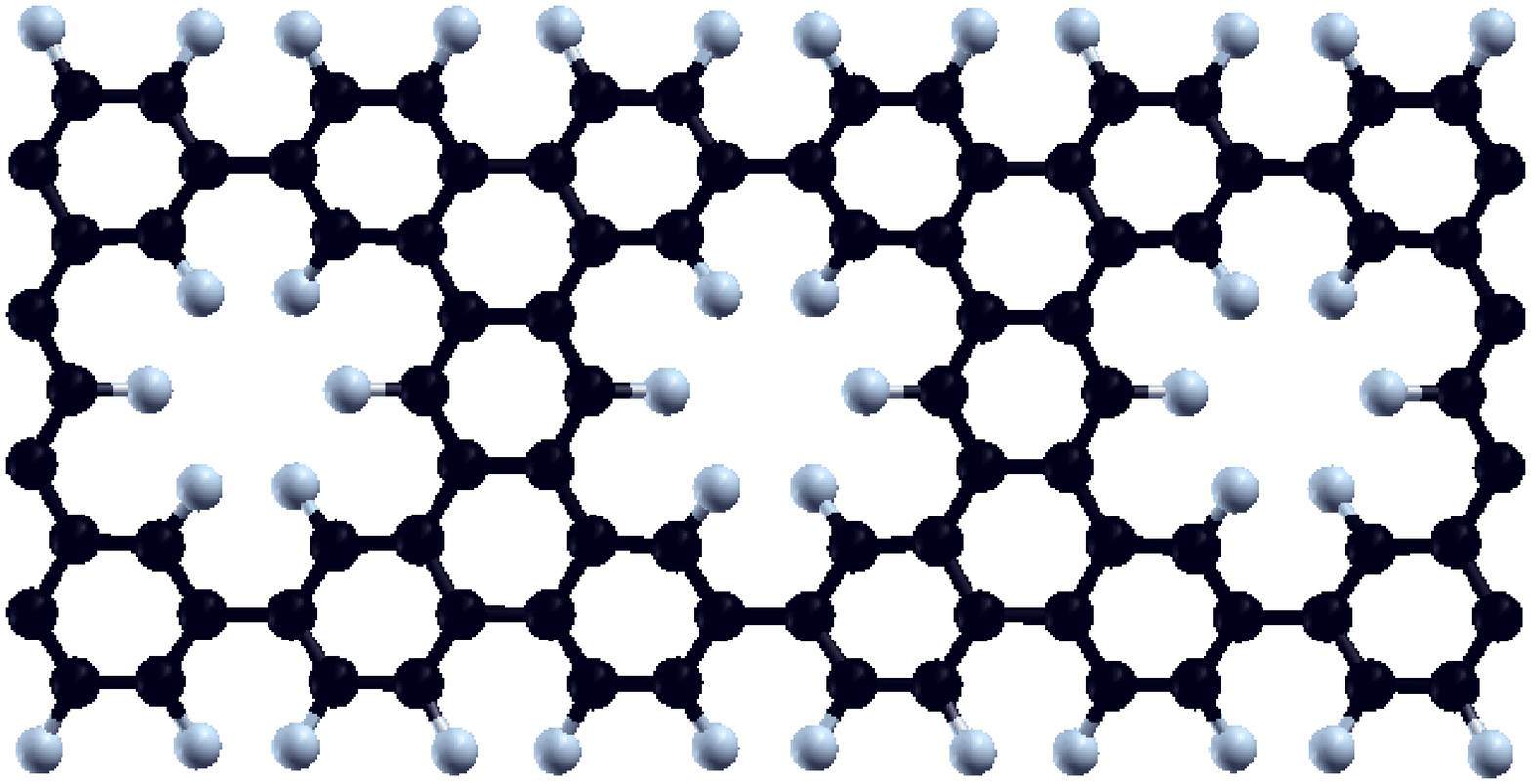}\hspace*{0.5cm}
\includegraphics[width=3.5cm]{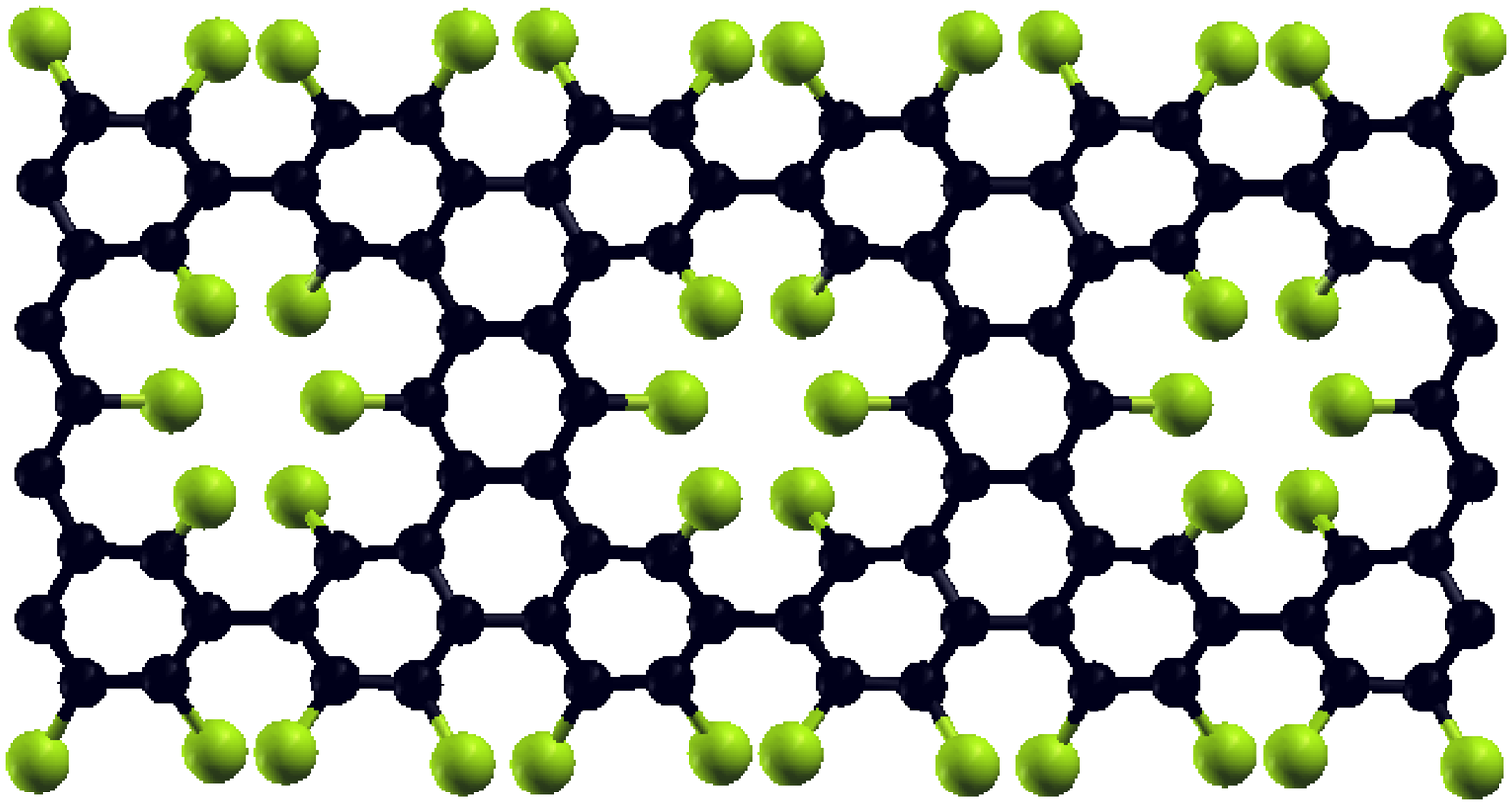}\vspace*{0.5cm}\\
\includegraphics[width=3.5cm]{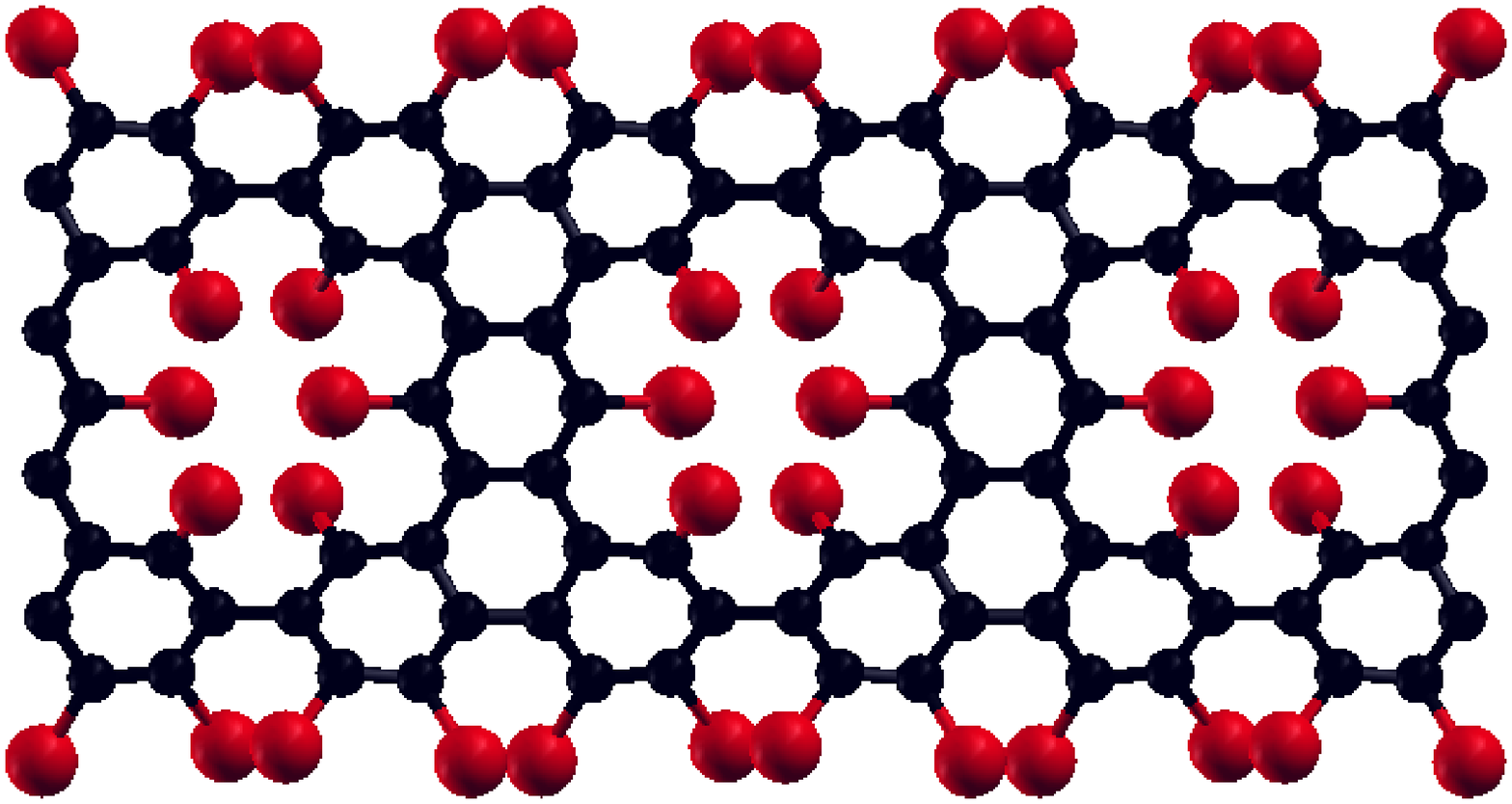}\hspace*{0.5cm}
\includegraphics[width=3.5cm]{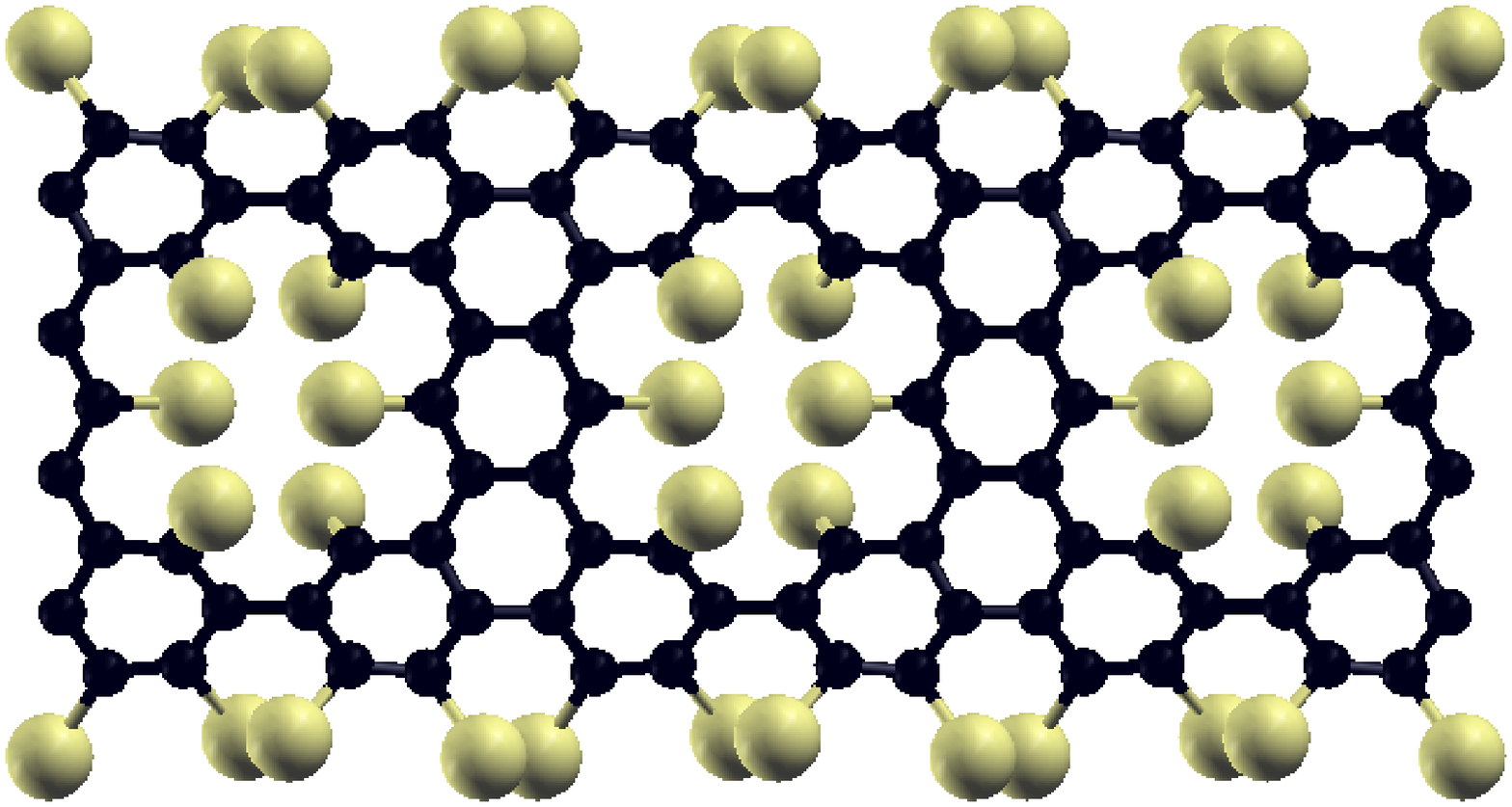}\vspace*{0.0cm}\\
(a) \vspace*{0.5cm}\\
\hspace*{-6cm}\vspace*{0.0cm}APG2\\
\includegraphics[width=6.5cm]{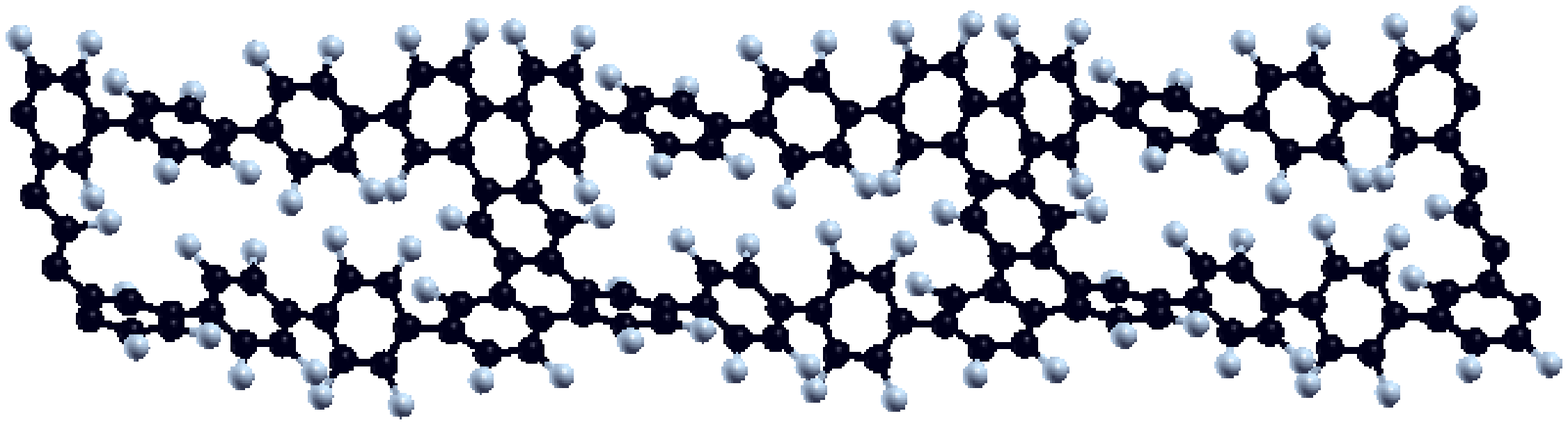}\vspace*{0.3cm}\\
\includegraphics[width=6.5cm]{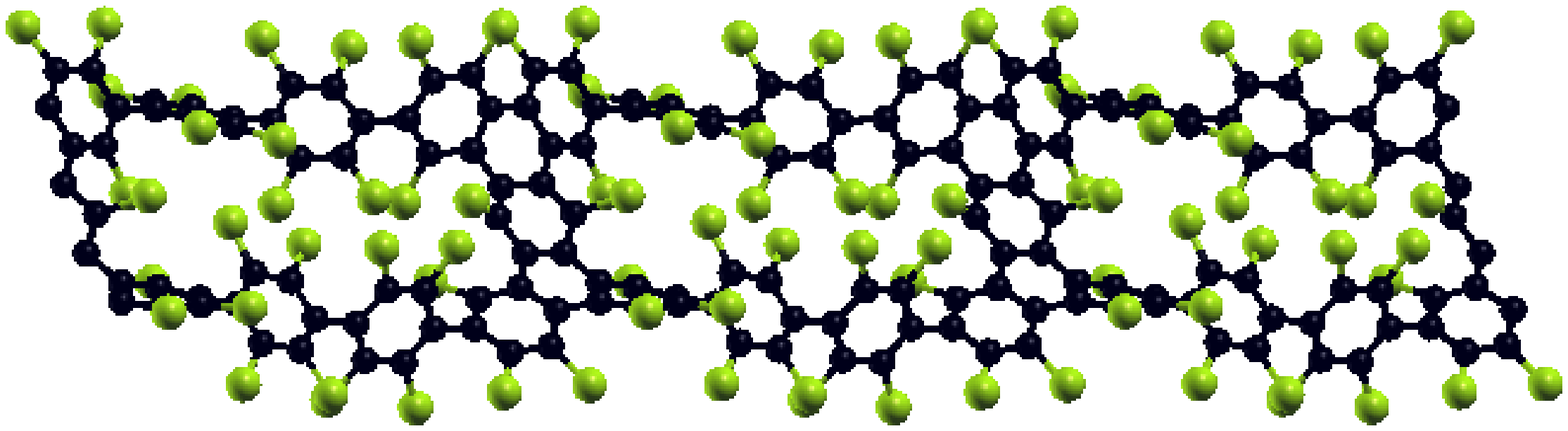}\vspace*{0.3cm}\\
\includegraphics[width=6.5cm]{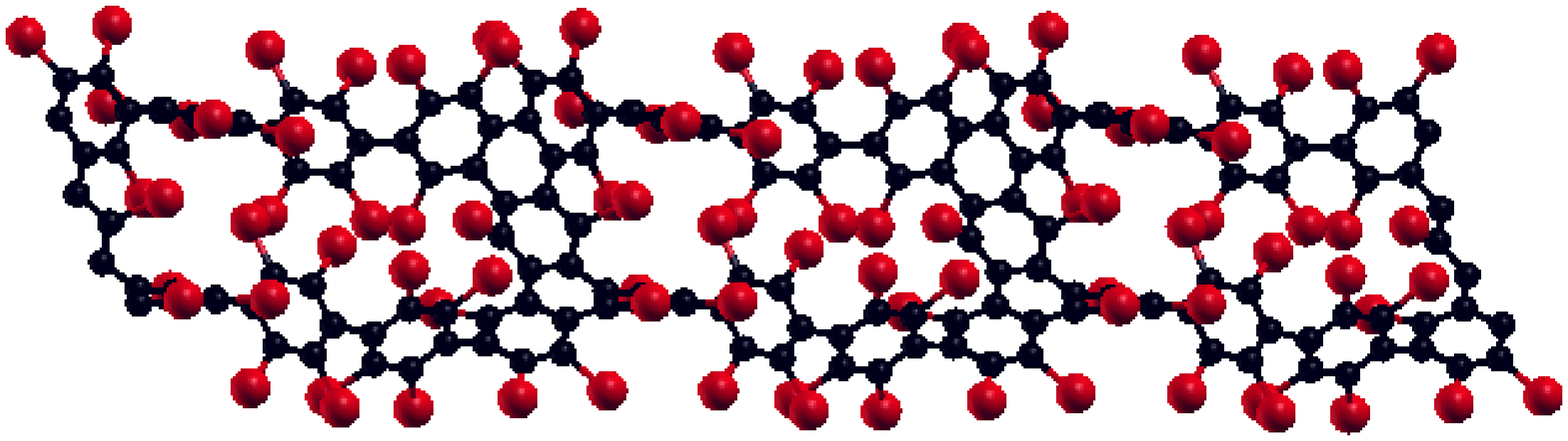}\vspace*{0.3cm}\\
\includegraphics[width=6.5cm]{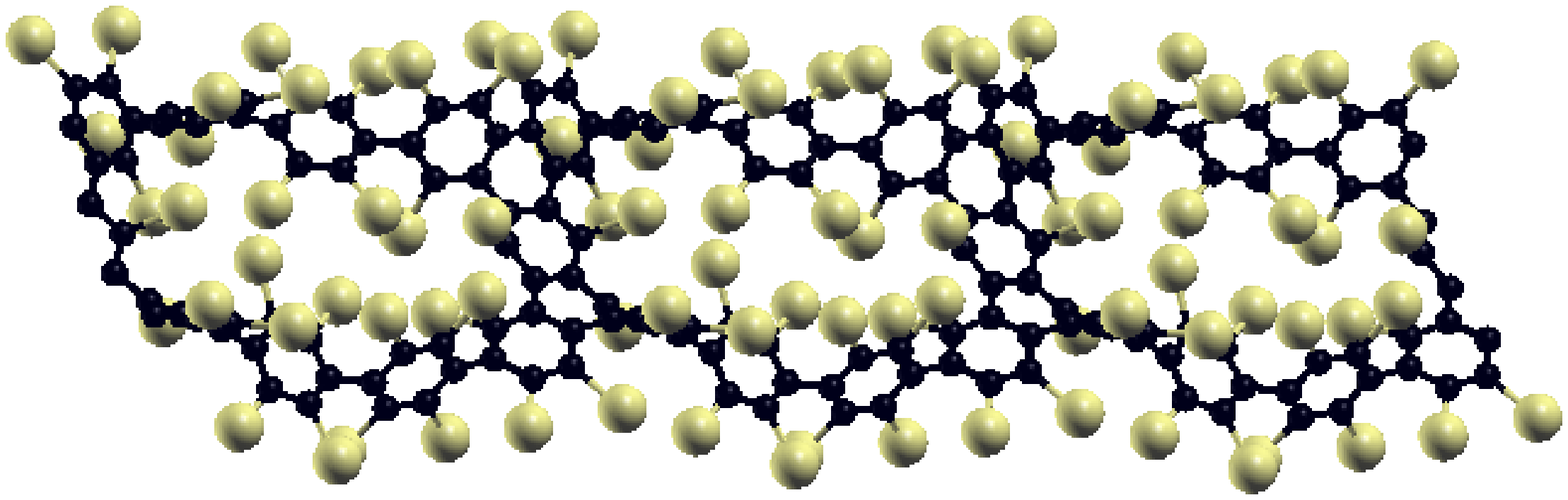}\\
(b)\\
\caption{Armchair nanoribbon structures APG1 (a) and APG2 (b),
passivated with different halogen atoms: F (blue), Cl (green), Br (red)
and I (yellow). Top and oblique views are shown for APG1 and APG2, 
respectively. Three unit cells are depicted in each case. 
The relative rotation of the passivated hexagons is enhanced as 
the atomic number of the halogen increases.
}
\label{apgstruct}
\end{figure}

In the present paper we investigate porous graphene nanoribbon structures passivated by halogen atoms (X = F, Cl, Br, I), as potential candidates for thermoelectric materials. The electronic band structures are determined for armchair and zig-zag terminated nanoribbons, while different sizes of the nanopores are considered for each type of structure. The influence of the halogen atoms on the electrical conduction is analyzed. The results are presented in relation with the vibrational properties in each case, which further give measure to the thermal conduction properties. Perspectives regarding the tunability of both electrical and thermal properties by passivation types and pore size are discussed.

\begin{figure}[t]
\centering
\hspace*{-6cm}\vspace*{0.0cm}ZPG1\\
\includegraphics[width=3.5cm]{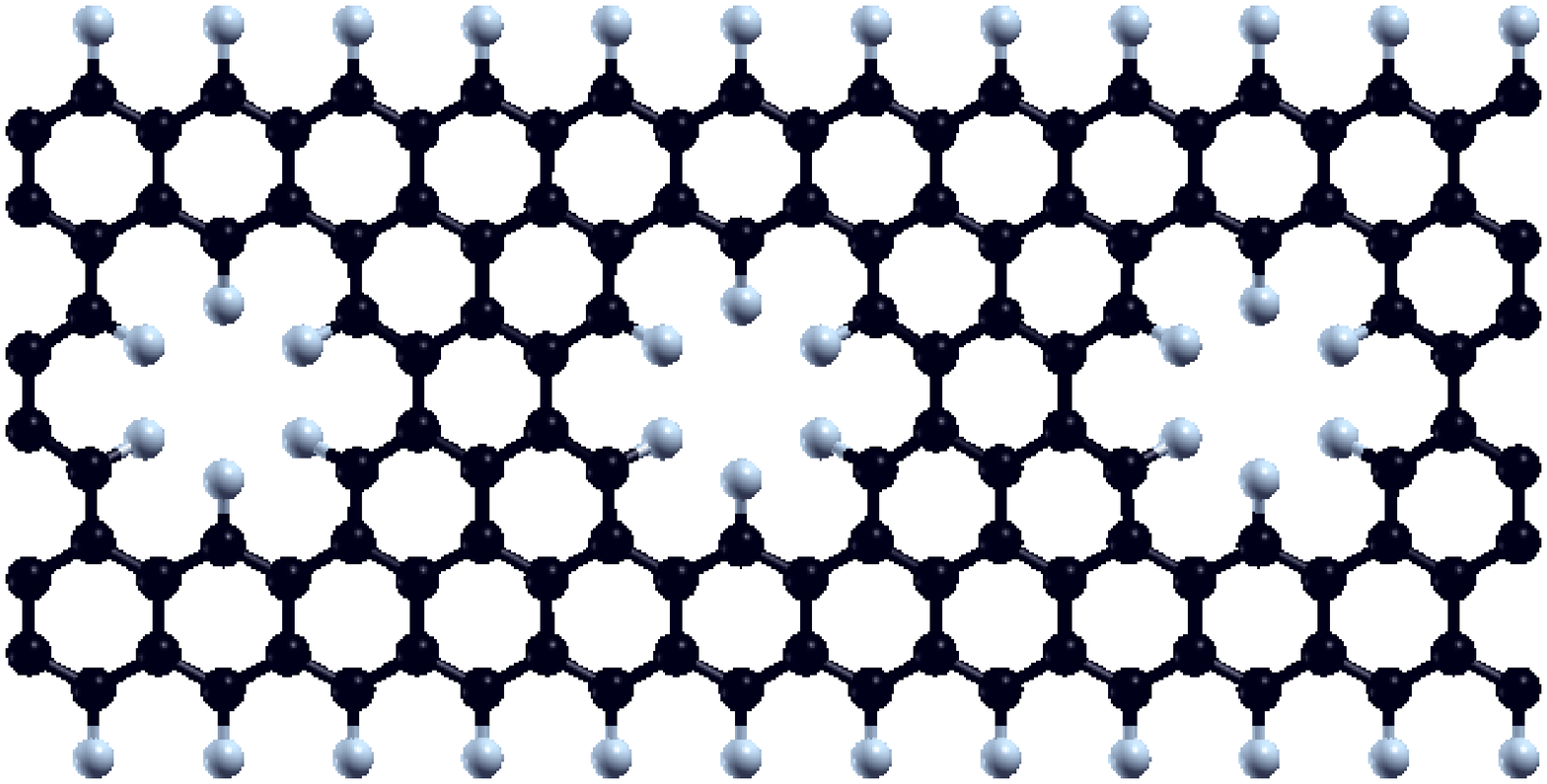}\hspace*{0.5cm}
\includegraphics[width=3.5cm]{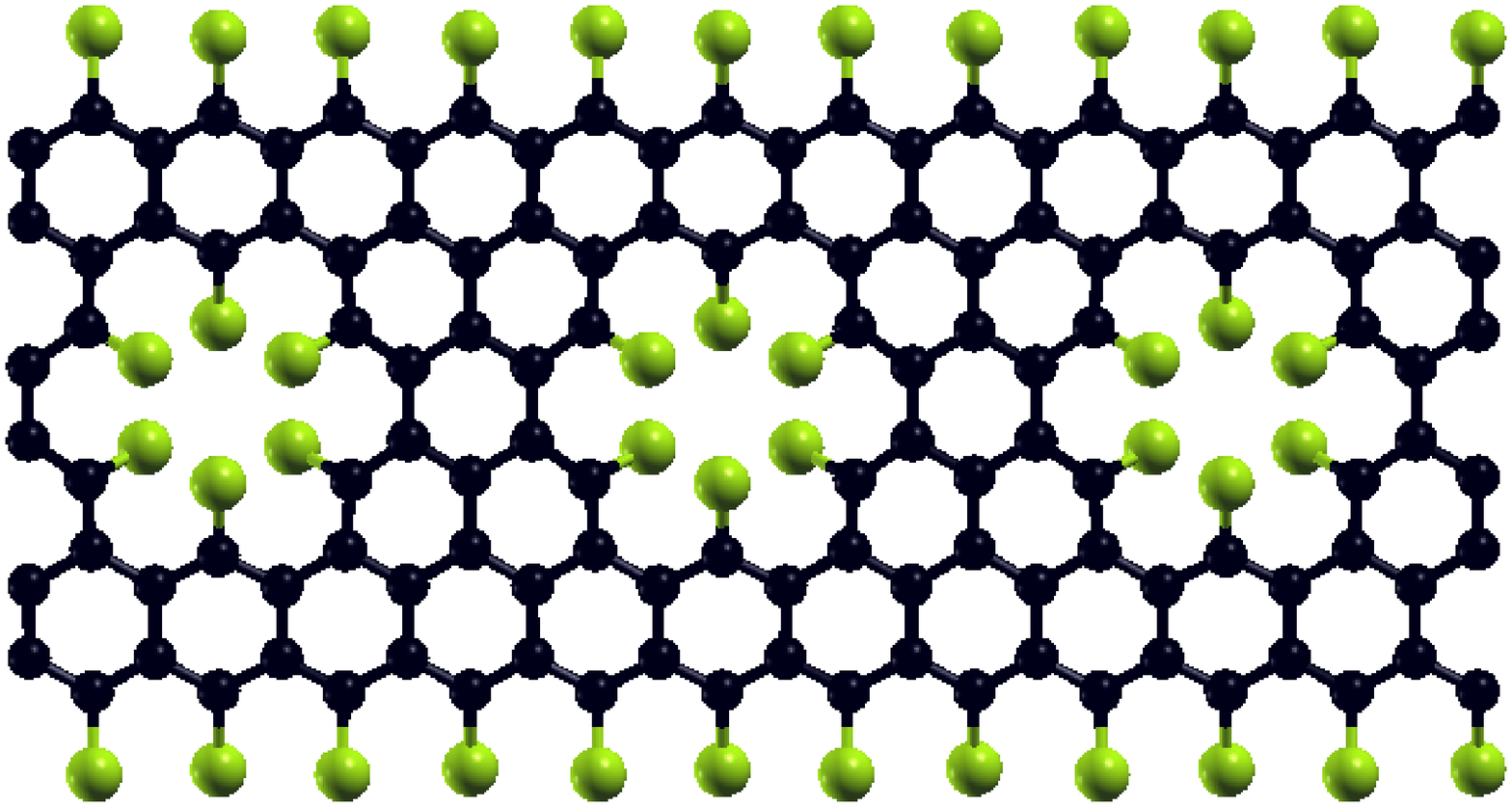}\vspace*{0.5cm}\\
\includegraphics[width=3.5cm]{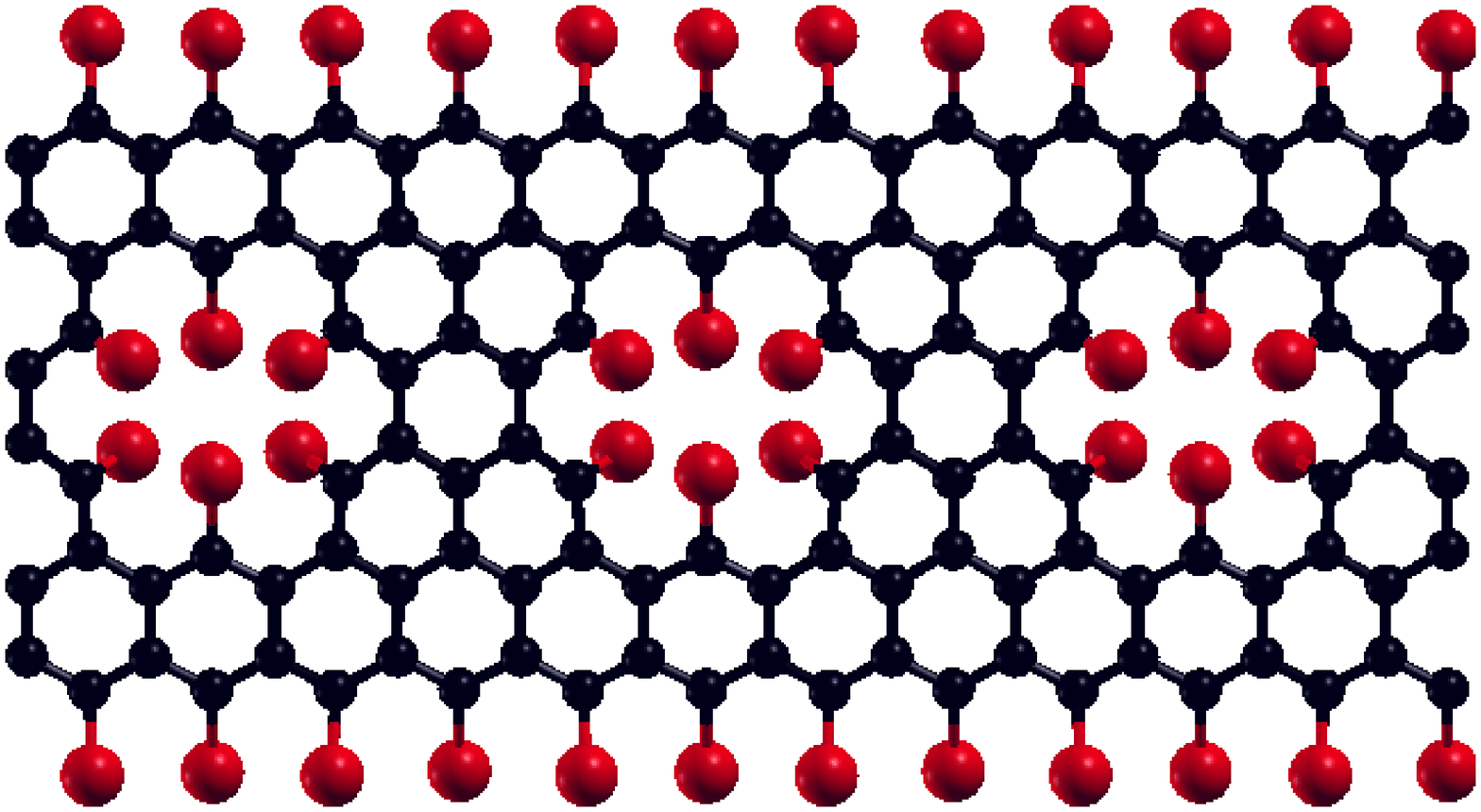}\hspace*{0.5cm}
\includegraphics[width=3.5cm]{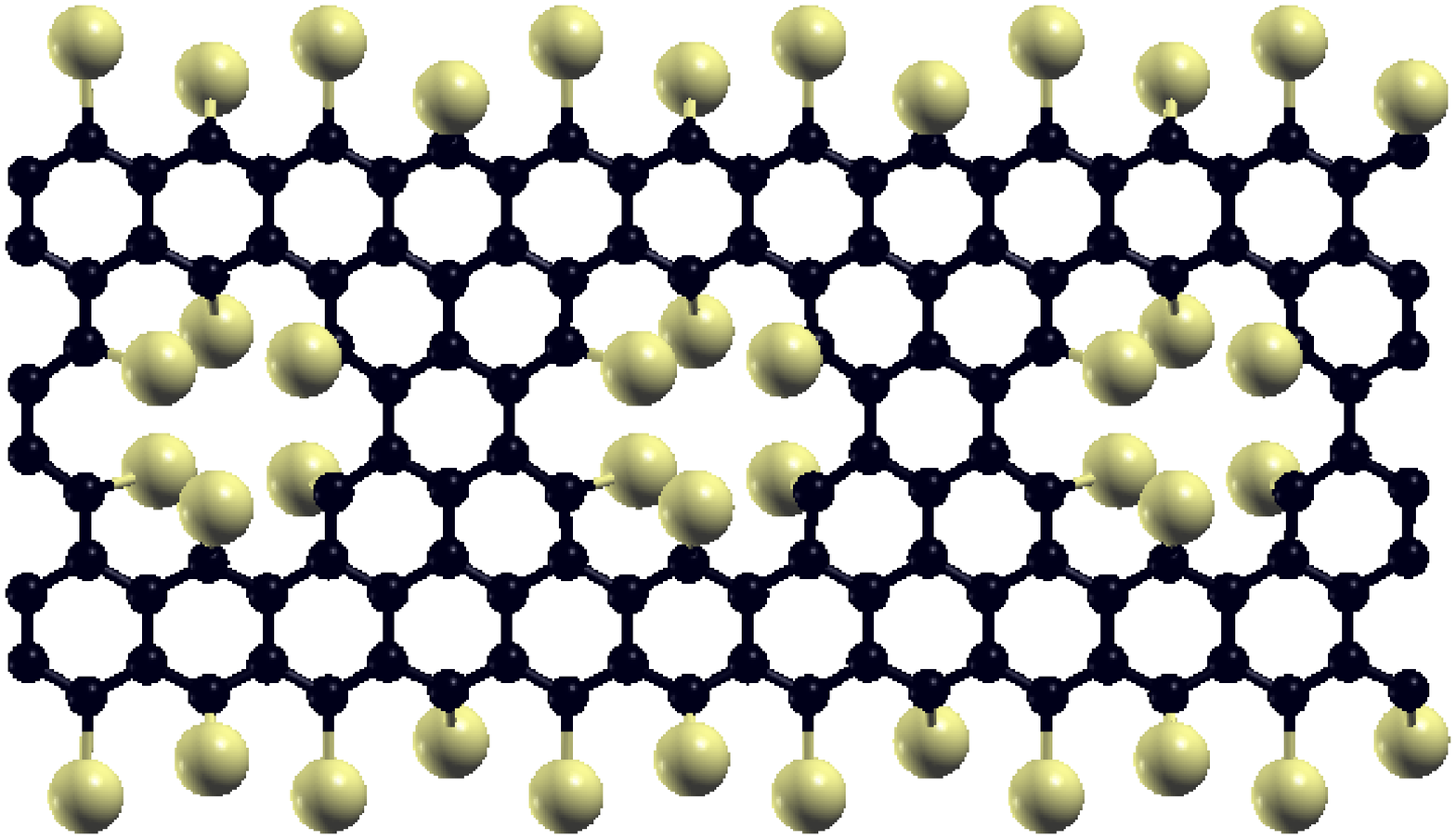}\vspace*{0.0cm}\\
(a) \vspace*{0.5cm}\\
\hspace*{-6cm}\vspace*{0.0cm}ZPG2\\
\includegraphics[width=6.5cm]{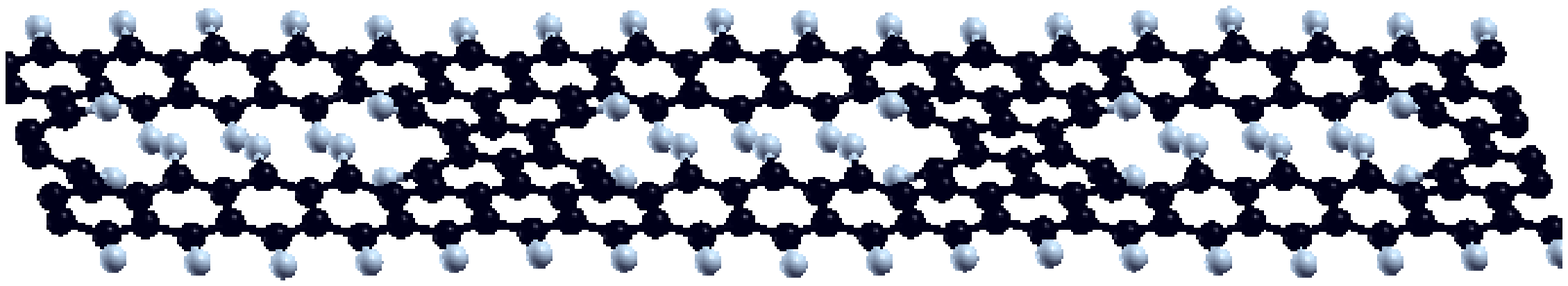}\vspace*{0.2cm}\\
\includegraphics[width=6.5cm]{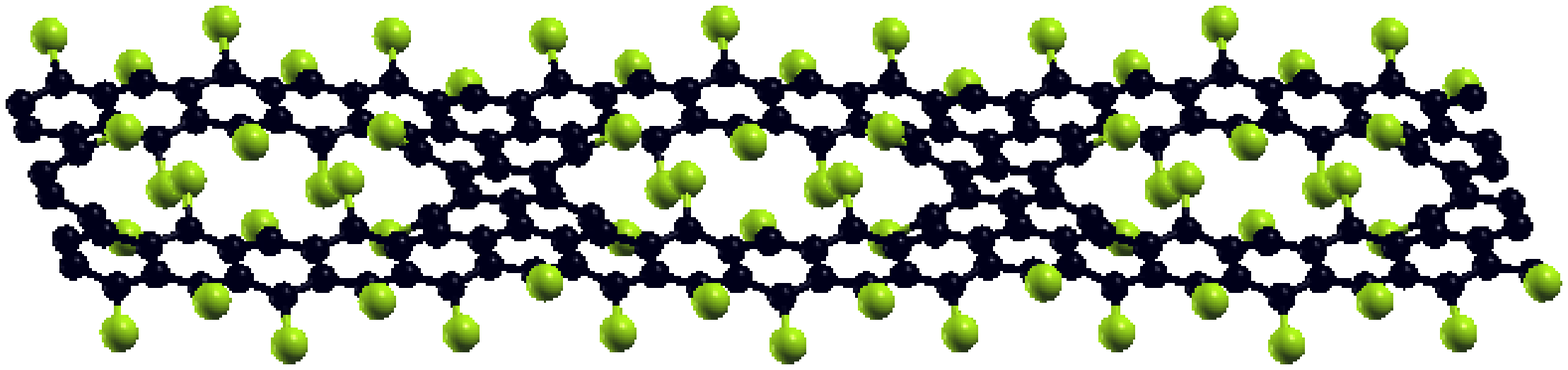}\vspace*{0.2cm}\\
\includegraphics[width=6.5cm]{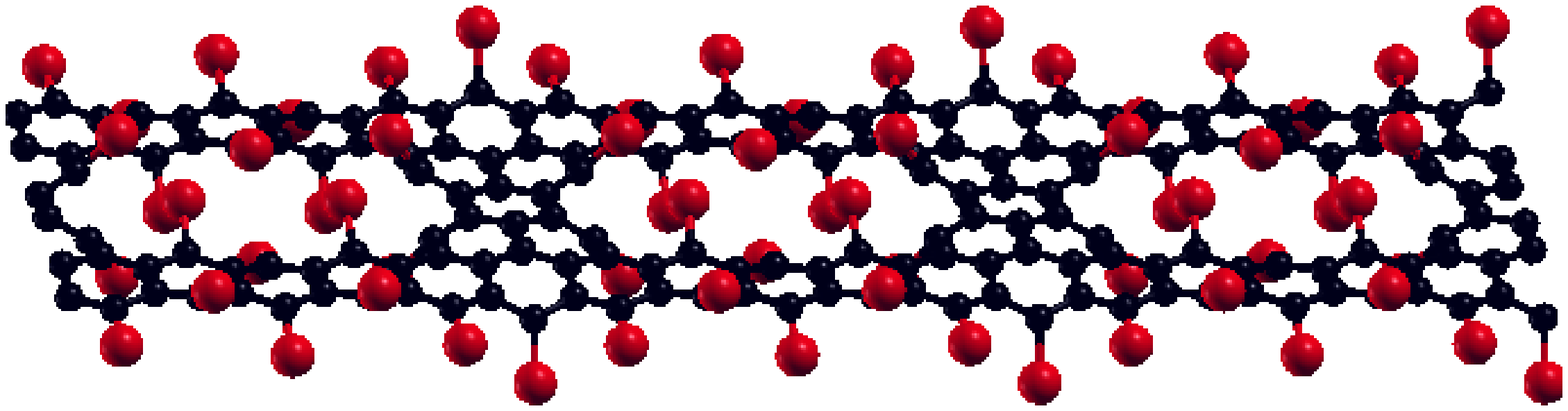}\vspace*{0.2cm}\\
\includegraphics[width=6.5cm]{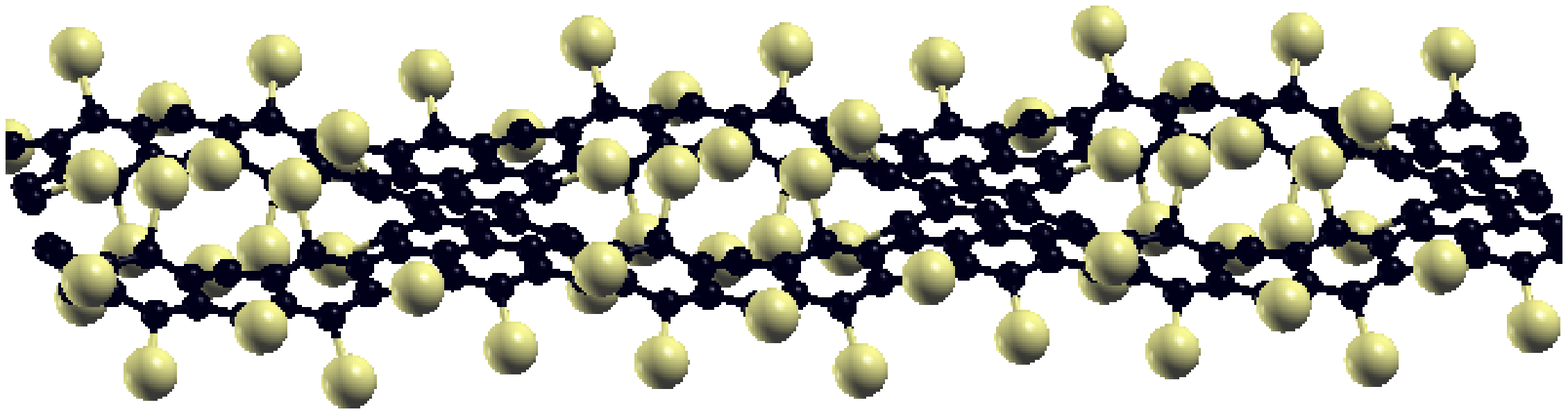}\\
(b)\\
\caption{Zigzag nanoribbon structures ZPG1 (a) and ZPG2 (b), with the same color codes as in Fig.\ \ref{apgstruct}.}
\label{zpgstruct}
\end{figure}

\section{Structures and methods}

The structures are graphene nanoribbons with periodically arranged pores, as depicted in Figs. \ref{apgstruct} and \ref{zpgstruct}. We label the structures according to the armchair and zig-zag terminations with APG and ZPG, respectively. For each termination type we consider two nanoribbon structures of the same width, but with different lengths of the unit cells, i.e. APG1/APG2 and ZPG1/ZPG2, which correspond to two sizes of the nanopores. The edges of the samples and the frontiers of the pores are halogen passivated. Thus we shall refer in the following to APG$\alpha$-X or ZPG$\alpha$-X, with $\alpha=1,2$ and X = F, Cl, Br, I.
The APG1/APG2 unit cells contain 30/54 carbon atoms and 14/30 halogen atoms, respectively, while ZPG1/ZPG2 have 42/58 carbon atoms and 14/22 halogen atoms.

Density functional theory (DFT) calculations were performed using SIESTA package \cite{soler}, which has the advantage of linear scaling of the computational time with the system size, achieved by using a strictly localized basis set. We employ the local density approximation (LDA) in the parameterization of Ceperley and Alder \cite{ceperley}. The double-zeta polarized (DZP) basis set was used, with the orbital-confining cutoff radii specified by an energy shift of 0.02 Ry and a standard split norm for the generation of multiple zeta in the split valence scheme of 0.15. Norm-conserving Troullier-Martins pseudopotentials were used. A mesh cut-off of 200 Ry was used for the real space discretization. The Monkhorst-Pack scheme employed in the sampling of
the Brillouin zone is $1\times1\times3$ k-points for the nanoribbon systems oriented along the $z$-direction.
The phononic band structures were calculated using the Vibra package, a module of SIESTA. The structural relaxations were performed until the residual forces were less than 0.01 eV/\AA. The force constants are determined using a $1\times1\times3$ supercell and displacing the atoms in the middle cell.

\begin{figure}[t]
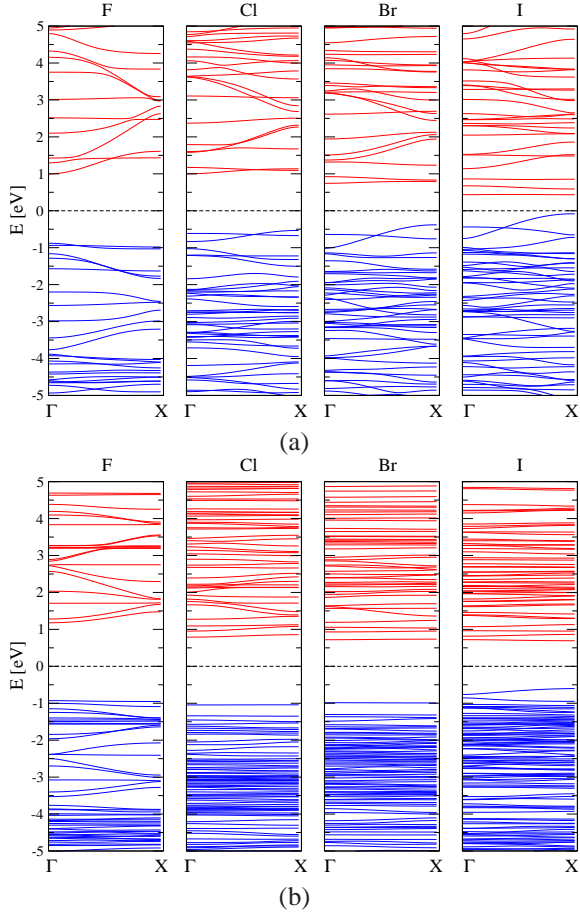

\centering
\includegraphics[width=7.5cm]{figure3a}\\
(a)\\
\includegraphics[width=7.5cm]{figure3b}\\
(b)
\caption{Electronic band structures of APG1 (a) and APG2 (b), showing semiconducting behavior. The dotted black lines mark the Fermi energy. The band gaps are decreasing with the atomic number of the halogen atom and are enhanced for larger pores. }
\label{el-bandsA}
\end{figure}

\section{Results and discussion}

Following relaxations, all computed structures converge to non-planar configurations observable in Figs. \ref{apgstruct} and \ref{zpgstruct}. The passivating halogen atoms are positioned along the sp$^2$ directions defined by the original graphene structure, although displaced out of the plane. In the case of the considered APG structures, the rotation allowed by the C-C bonds renders larger displacements of carbon atoms compared to the ZPG structures, where the elementary hexagons containing six carbon atoms, present along the edge and between the pores, enhance the rigidity. APG2 structures have four C-C bonds per unit cell along each edge, which allow the relative rotation of the hexagons, compared to only two C-C bonds in APG1 structures. The out of plane displacement of the halogen atoms increases with the atomic number. In the minimum energy configurations of the ZPG structures the halogen atoms are alternating above and below the nanoribbon plane at the edges and also inside the nanopores. In the case of APG structures, groups of two adjacent halogen atoms are jointly displaced in the same direction inducing the elementary hexagon rotations. 
The fluorine passivated ZPG structures, ZPG1-F and ZPG2-F, present the smallest deformations, while at the opposite end the APG2-I structure is found.

\begin{figure}[t]
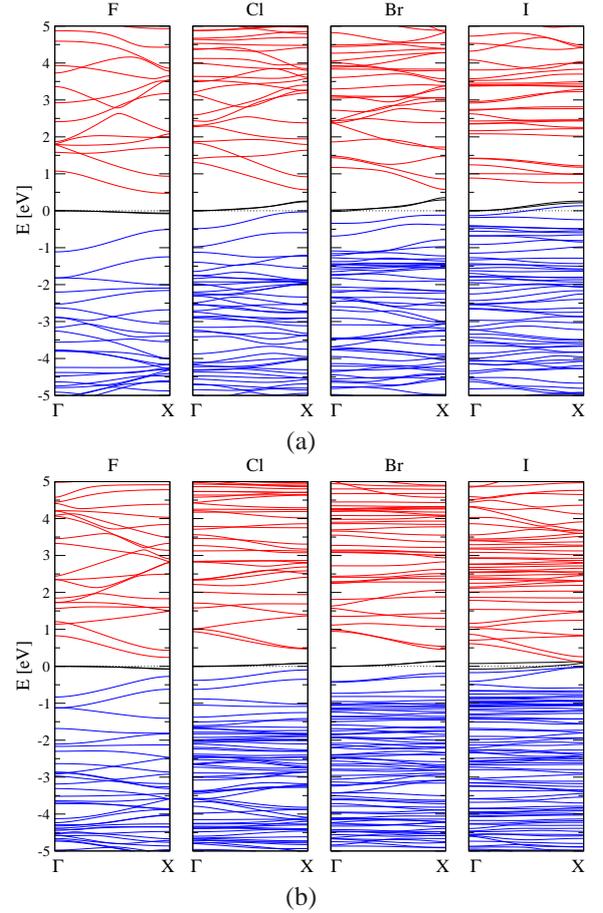

\centering
\includegraphics[width=7.5cm]{figure4a}\\
(a)\\
\includegraphics[width=7.5cm]{figure4b}\\
(b)
\caption{Electronic band structures of metallic ZPG1 (a) and ZPG2 (b) nanoribbons. The bands crossing the Fermi energy are depicted in black.}
\label{el-bandsZ}
\end{figure}

Electronic band structures of APG and ZPG porous nanostructures depicted in Figs.\ \ref{el-bandsA} and \ref{el-bandsZ} show similarities to the pristine graphene nanoribbon counterparts: while APG structures are semiconducting for narrow widths, the ZPG structures exhibit metallic behavior irrespective of their lateral sizes. The choice of the passivating halogen however influences the magnitude of the gap in the case of APG structures, as indicated in Table\ \ref{tabone}. For larger atomic numbers, the overlap of the orbitals diminishes the band gaps, which are still visible for APG1-I and APG2-I systems.

\begin{figure}[t]
\centering
\includegraphics[width=7.8cm]{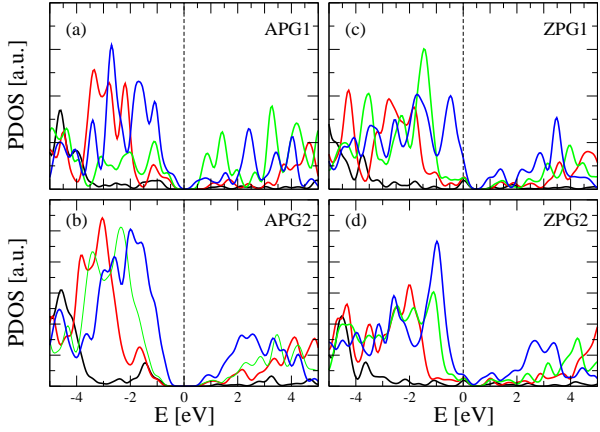}\\
\caption{Partial density of states corresponding to contributions from different halogens:
F (black), Cl (red), Br (green), I (blue). The vertical dashed lines mark the Fermi level.}
\label{pdos}
\end{figure}

The band gaps are decreasing from 1.87 eV (APG1-F) to 0.87 eV (APG1-I) and from 2.11 eV (APG2-F) to 1.48 eV (APG2-I). This indicates the feasibility of band gap tuning by halogen passivation. On the other hand, all ZPG structures show an increasingly higher density of states at the Fermi energy by switching from fluorine to iodine as passivating agent. The same overall behavior was found previously in halogenated nanoribbons without pores, in either pristine graphene nanoribbons, in boron-nitride nanoribbons or mixtures of both \cite{nemnes2}.

\begin{table}[h]
\begin{center}
\begin{tabular}{*{6}{l}}
\hline
\hline                            
  & \hspace*{0.8cm}F & \hspace*{0.8cm}Cl & \hspace*{0.8cm}Br & \hspace*{0.8cm}I \cr 
\hline
APG1 & \hspace*{0.8cm}1.87 & \hspace*{0.8cm}1.61 & \hspace*{0.8cm}1.38 & \hspace*{0.8cm}0.87  \cr
APG2 & \hspace*{0.8cm}2.11 & \hspace*{0.8cm}1.83 & \hspace*{0.8cm}1.71 & \hspace*{0.8cm}1.48  \cr 
\hline
\hline
\end{tabular}
\end{center}
\caption{\label{tabone}Energy band gaps [eV] of APG structures for the two nanopore sizes considered, i.e. APG1 and APG2. The ZPG structures are gapless.} 
\end{table}

The diminished band gaps of the APG structures are correlated with the partial density of states (PDOS), as described in Fig.\ \ref{pdos}. For all structures the contributions of the halogens become significantly important near the Fermi energy, in the F-Cl-Br-I sequence. More specifically, the proportion to the total density of states of  $2p^5$ F, $3p^5$ Cl, $4p^5$ Br, $5p^5$ I orbitals is the largest for each structure, but there are also important contributions from hybridized $2p^2$ orbitals of carbon. As discussed before, the distortions in the nanoribbons are minimal for fluorine passivation compared to the structures containing the other three halogens. This effect is captured also in Fig.\ \ref{pdos}, where the fluorine PDOS is peaked at comparably smaller energy values.

\begin{figure}[t]
\centering
\includegraphics[width=7.8cm]{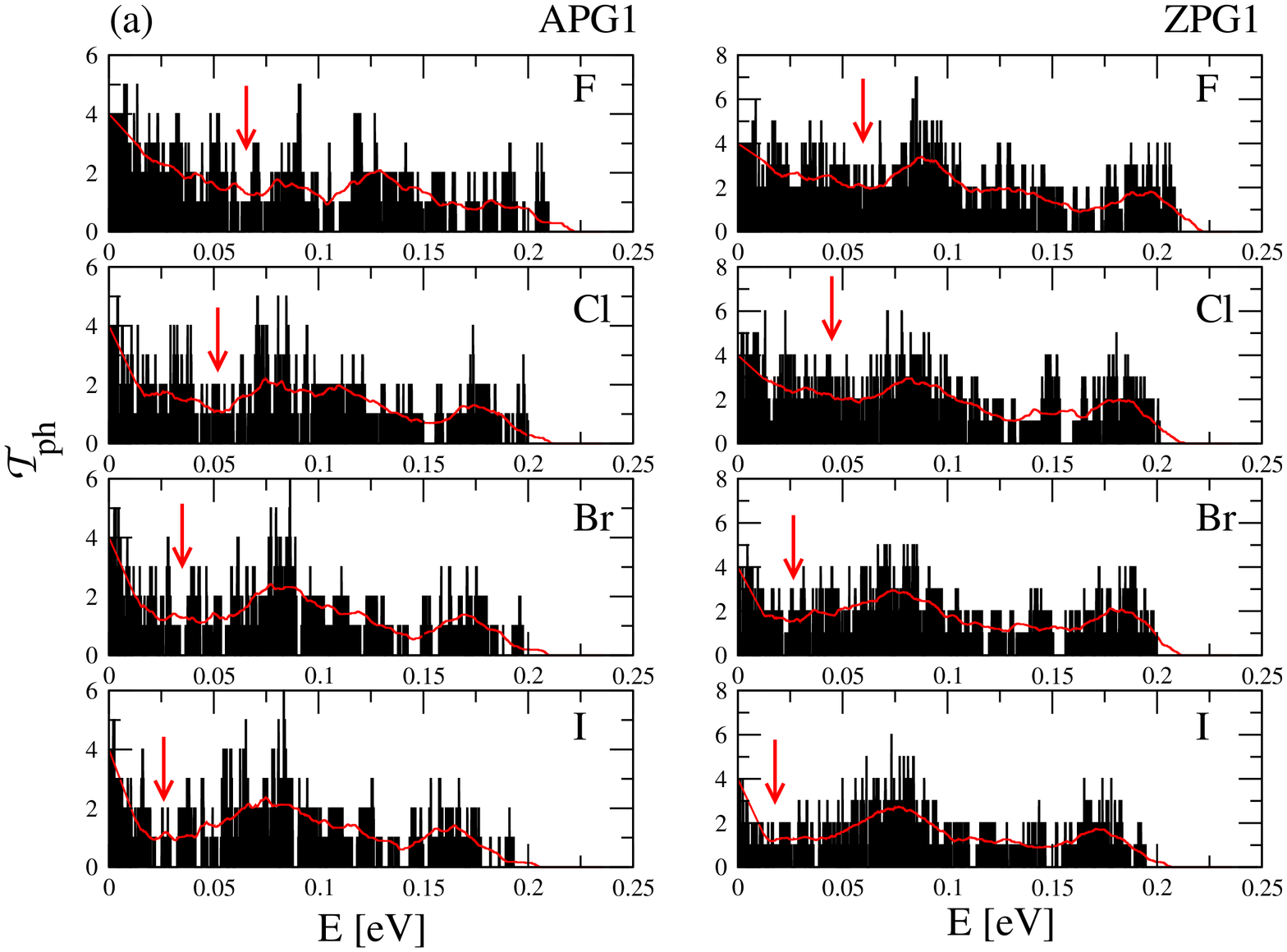}\\
\includegraphics[width=7.8cm]{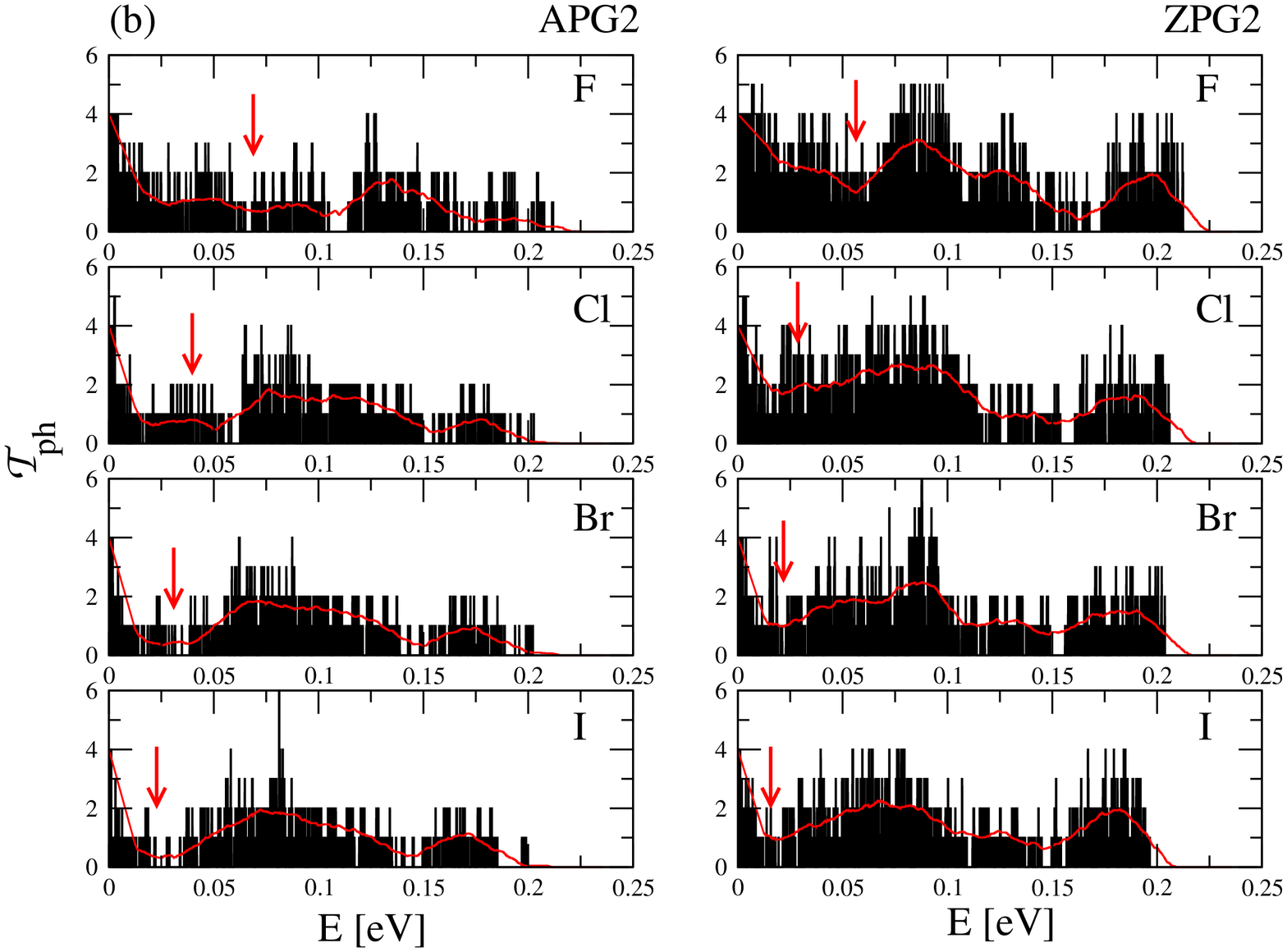}\\
\caption{Phononic (${\mathcal T}_{ph}$) transmission functions
for the armchair and zig-zag nanoribbons, APG1/ZPG1 (a) and APG2/ZPG2 (b). The red solid lines represent averages of ${\mathcal T}_{ph}$. The arrows mark the positions of the pseudo-gaps, which are shifted towards lower energies in the F-Cl-Br-I sequence.}
\label{ph_trans}
\end{figure}

The thermal conduction properties rely, in addition to the electronic component, on the phononic transmission functions, ${\mathcal T}_{ph}$, which are shown in Fig.\ \ref{ph_trans}.
The ${\mathcal T}_{ph}$ functions present a systematic shift of the spectrum towards lower frequencies as the mass of halogen atoms increases. However the overall distribution of phonon modes within the band structures is preserved in each case. Of interest for the thermal conductance are the presence of pseudo-gaps located in the lower part of the spectrum, which corresponds to relatively low thermal conductivity around the room temperature. The origin of the pseudo-gaps is related to the periodic distribution of pores, in contrast to pristine graphene nanoribbons \cite{nemnes}. Their position is shifted towards lower energies following the change from F to I. At the same time the minima indicated by the arrows become more pronounced. By increasing the size of the nanopore these effects are further enhanced.

\begin{figure}[t]
\centering
\includegraphics[width=7.8cm]{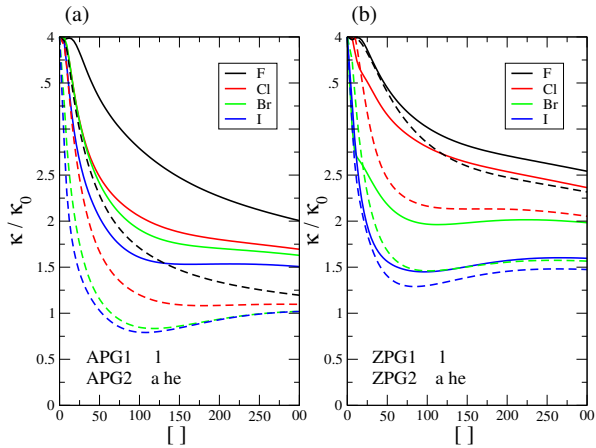}\\
\caption{Thermal conductance of APG (a) and ZPG (b) structures passivated with halogens, for the two sizes of the nanopores considered, i.e. APG1/ZPG1 (solid lines) and APG2/ZPG2 (dashed lines). 
}
\label{kappa}
\end{figure}

Using the phononic transmission functions we determine the thermal conductance, calculated according to the relation \cite{markussen,nemnes3}:
\begin{equation}
\hspace*{-0.6cm}
\kappa_{ph}(T) = \frac{\hbar^2}{2\pi k_{\mbox{\scriptsize B}} T^2} 
	\int_0^\infty d\omega \; \omega^2 \; {\mathcal T}_{\mbox{\scriptsize ph}}(\omega) 
           \frac{\exp^{\hbar\omega/k_{\mbox{\scriptsize B}}T}}
                {(\exp^{\hbar\omega/k_{\mbox{\scriptsize B}}T}-1)^2}
\label{kappa-eq}
\end{equation}
where $k_{\mbox{\scriptsize B}}$ is the Boltzmann constant. 
Fig. \ref{kappa} shows a systematic decrease of the scaled thermal conductance, for all structures, as the halogen atomic number is increased. This effect is further amplified in each case by enlarging the nanopore size. The scaling factor is the thermal conductance quantum $\kappa_0(T)=\pi^2 k_{\mbox{\scriptsize B}}^2 T / 3h$.
Finite width nanoribbons are quasi one-dimensional systems and there are four acoustic phonon branches: the out-of plane mode (ZA), the in-plane transverse mode (TA), the in-plane longitudinal (LA) and a torsion branch, which does not exist for an infinite width, i.e. for the two-dimensional graphene sheet \cite{wang,gilen,zhang2}.

Furthermore, comparing the APG with the ZPG structures, one can see larger
differences in $\kappa_{ph}(T)$ for the two nanopore sizes in the former
case, which may develop from an enhanced deformation of the APG
structures. In addition, the thermal conductance of Br and I
passivated structures have minima in the temperature interval 50-100
K, resulting from the well defined minima in the ${\mathcal T}_{ph}$
functions, found at energies lower than for the other two
structures (Fig. \ref{ph_trans}). For temperatures smaller than 100
K the dominant effect is the decrease of ${\mathcal T}_{ph}$, while for
larger temperatures the phonon modes become significantly populated 
above the minima of the phononic transmission function, resulting in
a slight increase of $\kappa_{ph}$.

The decrease of the scaled thermal conductance is qualitatively similar with the one obtained in the case of two dimensional polycyclic carbon networks \cite{nemnes}. The thermal conductance is also lower than for the pristine zig-zag graphene nanoribbons passivated with hydrogen \cite{huang,nemnes}. Similar phonon transmission spectra, with pseudo-gaps present at low frequencies, were obtained by nano-structuring the graphene nanoribbons with periodic patterns \cite{sevincli2}. The tunability of both electronic and thermal properties, already pointed out in a recent experimental study \cite{poh}, is here demonstrated for narrow widths nanoribbons and deserves further investigations.

\section{Conclusions}

Porous graphene nanoribbons passivated with halogens have been investigated by {\it ab initio} DFT calculations. It is shown that the overall electrical behavior of armchair and zig-zag terminated nanoribbons is preserved, i.e. semiconducting and metallic behaviors, respectively. However, as the atomic number of the halogen increases smaller band gaps are observed for APG structures and a higher density of states at the Fermi level is achieved for the ZPG structures. The phononic transmission functions have minima corresponding to pseudo-gaps which shift to the low energy domain of 50-100 K by switching from fluorine to iodine passivation. 
Increasing the size of the nanopores, the pseudo-gaps become better defined and consequently the thermal conductance decreases, exhibiting minima for bromine and iodine passivated systems, due to the pseudo-gaps in the phononic spectrum found at lower energies. The decrease in  the scaled thermal conductivity is in contrast to the pristine graphene case.
These observations indicate the possibility of tuning both the electronic and thermal properties by halogen passivation, i.e. enhancing the electrical conduction and, at the same time, lowering the thermal conductance, which is a goal for achieving efficient thermoelectric devices.\\

{\bf Acknowledgements}\\

Discussions with Dr. Dragos-Victor Anghel are highly appreciated.  
This work was supported by the National Authority for Scientific Research and Innovation
(ANCSI) under grant PN16420202.


\section*{References}





\end{document}